\documentclass[floatfix,aps,prl,groupedaddress, twocolumn]{revtex4-2}
\usepackage{braket}
\usepackage{amsmath}
\usepackage{svg}
\usepackage[utf8]{inputenc}
\usepackage[T1]{fontenc}
\usepackage{lineno}
\usepackage{comment}
\usepackage{graphicx}
\usepackage[normalem]{ulem}
\usepackage[colorlinks,breaklinks]{hyperref}
\definecolor{darkred}{rgb}{0.5,0,0}
\definecolor{darkgreen}{rgb}{0,0.5,0}
\definecolor{darkblue}{rgb}{0,0,0.5}
\hypersetup{linkcolor=darkred,citecolor=darkgreen,urlcolor=darkblue}

\usepackage[toc,page]{appendix}
\usepackage{tikz}
\usetikzlibrary{math}
\usetikzlibrary{decorations.pathreplacing,calligraphy}

\begin{document}

\title{Simultaneously sorting overlapping quantum states of light}

\author{Suraj Goel}
\thanks{These authors contributed equally}

\affiliation{School of Engineering and Physical Sciences, Heriot-Watt University, Edinburgh, EH14 4AS, UK}

\author{Max Tyler}
\thanks{These authors contributed equally}

\affiliation{School of Engineering and Physical Sciences, Heriot-Watt University, Edinburgh, EH14 4AS, UK}

\author{Feng Zhu}

\affiliation{School of Engineering and Physical Sciences, Heriot-Watt University, Edinburgh, EH14 4AS, UK}

\author{Saroch~Leedumrongwatthanakun}

\affiliation{School of Engineering and Physical Sciences, Heriot-Watt University, Edinburgh, EH14 4AS, UK}

\author{Mehul Malik}

\affiliation{School of Engineering and Physical Sciences, Heriot-Watt University, Edinburgh, EH14 4AS, UK}

\author{Jonathan Leach}
\email{j.leach@hw.ac.uk}

\affiliation{School of Engineering and Physical Sciences, Heriot-Watt University, Edinburgh, EH14 4AS, UK}

\begin{abstract}
The efficient manipulation, sorting, and measurement of optical modes and single-photon states is fundamental to classical and quantum science.  Here, we realise simultaneous and efficient sorting of non-orthogonal, overlapping states of light, encoded in the transverse spatial degree of freedom.  We use a specifically designed multi-plane light converter (MPLC) to sort states encoded in dimensions ranging from $d = 3$ to $d = 7$. Through the use of an auxiliary output mode, the MPLC simultaneously performs the unitary operation required for unambiguous discrimination and the basis change for the outcomes to be spatially separated. Our results lay the groundwork for optimal image identification and classification via optical networks, with potential applications ranging from self-driving cars to quantum communication systems.

\end{abstract}

\maketitle


 \emph{Introduction}:-- The task of discriminating between a set of quantum states is a fundamental requirement in quantum information science, and in particular, quantum communication \cite{Helstrom_1969, nielsen_chuang_2010}. However, in general, two different quantum states can have a finite, non-zero overlap with respect to each other, making them non-orthogonal and therefore difficult to separate. It is theoretically impossible to perform a measurement that allows us to perfectly distinguish between such states 100\% of the time. An important question now follows: given a set of quantum states with a non-zero overlap, what is the best measurement strategy to distinguish between them? 

\begin{figure}[t!]
\centering
\includegraphics[width=3.3in]{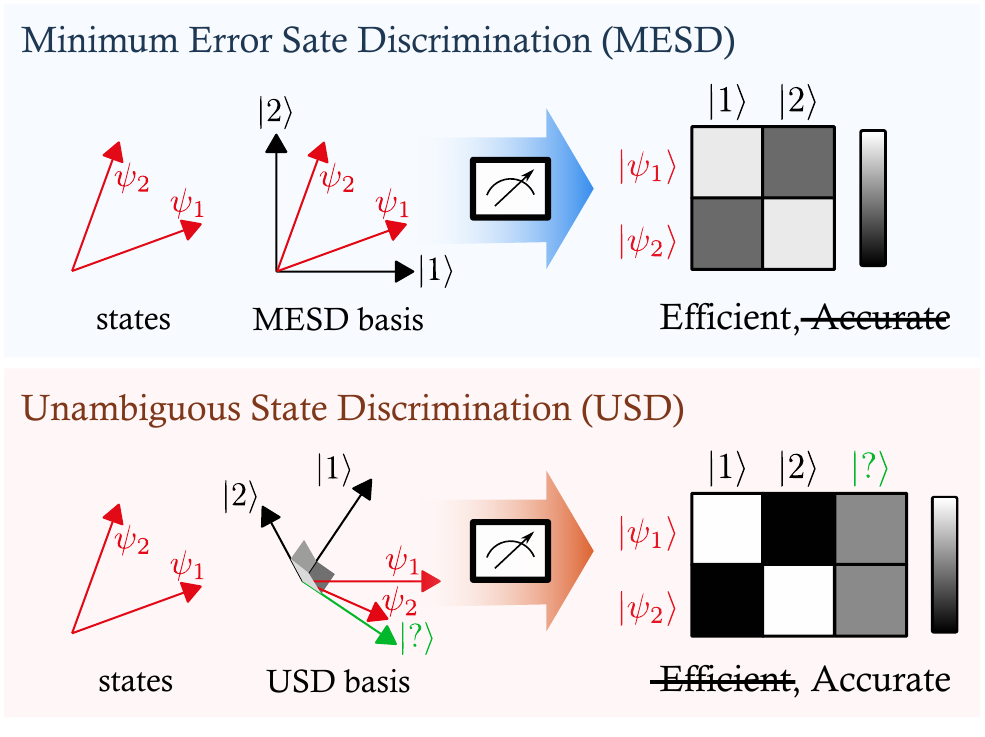}
\caption{Measuring non-orthogonal states using minimum error or unambiguous state discrimination. The MESD protocol is efficient in that every input state is always categorised, even if this leads to errors.  The USD protocol is accurate in that every input state is correctly identified, even if this does not happen 100\% of the time.  } \label{Fig:concept}
\end{figure}

The answer to this question lies in quantum measurement theory, where strategies for the optimal measurement of non-orthogonal quantum states are known~\cite{Barnett2009, chefles00_quant_state_discr}.  In the extreme, we are left with a choice between a measurement strategy that is either efficient OR accurate (only orthogonal states can be sorted efficiently AND accurately), see Fig.~\ref{Fig:concept}. The efficient option is minimum error state discrimination (MESD) \cite{Barnett_2009}. Here, one seeks to perform a set of measurements that categorise every input state.  The drawback to MESD is that errors are inevitable, and we have to accept that we will be incorrect with some probability relating to the overlap of the input states.  The accurate option is unambiguous state discrimination (USD) \cite{Ivanovic1987, CHEFLES1998223, CHEFLES1998339, Clarke2001, franke-arnold12_unamb_state_discr_high_dimen}. Here, one seeks to perform measurements that never incorrectly identify the input state.  The downside here is that state identification occurs with a reduced probability, i.e.~a measurement does not always provide a result, but when it does, it is always correct.    In addition to the extremes, there are intermediate strategies that have been developed that interpolate between MESD and USD.  Several theoretical studies include discrimination with a known error margin \cite{Hayashi2008, Sugimoto2009, Sugimoto2012, Sentis2013}, discrimination with a fixed rate of inconclusive outcomes \cite{Bagan2012, Herzog2012, Jimenez2021}, and partial state separation with a MESD procedure \cite{Nakahira2012,Nakahira2015a,Nakahira2015b, Solis2016,Nakahira2017}.

  The problem that we address in this work is practical high-dimensional unambiguous state discrimination, i.e.~a positive operator-valued measure (POVM) for non-orthogonal, high-dimensional states with simultaneous outcomes.  High-dimensional quantum states (or qudits) allow for quantum information to be encoded in a $d$-dimensional space, enabling quantum communication protocols with increased information capacity and robustness to noise~\cite{Mirhosseini_2015,Pivoluska_2018,zhu19_are_high_dimen_entan_states, PhysRevX.9.041042, srivastav2022noise}.  While the theoretical foundation for the necessary measurement strategies has already been developed, their experimental realisation has proved to be a significant challenge.

\begin{figure*}[htp]
\centering
\includegraphics[width=17.9cm]{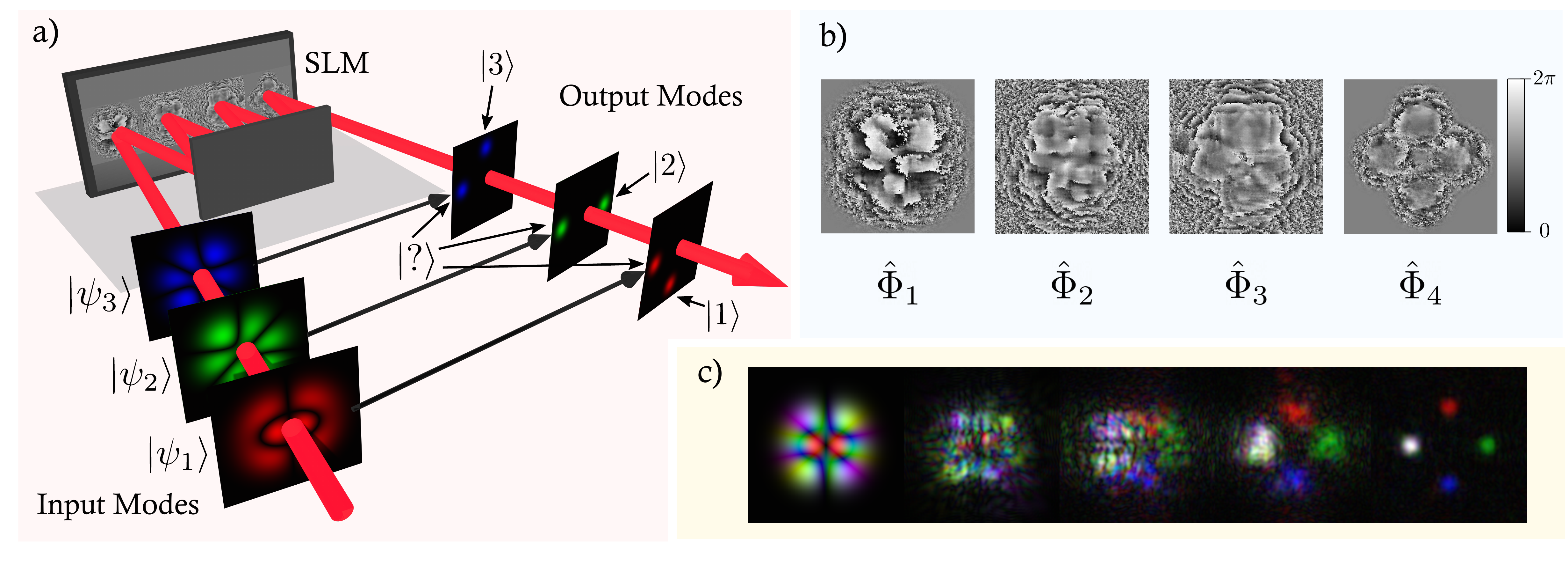}\caption{(a) Schematic of the mode sorter for non-orthogonal states with colour used to represent different input modes.  The $d$ input modes pass through a multi-plane light converter that sorts them into $d+1$ orthogonal outputs of spatially separated Gaussian spots.  The one additional output mode $\ket{?}$ corresponds to the ambiguous outcome. (b) Example holograms used for the MPLC. (c) Amplitudes of non-orthogonal modes as they propagate through the MPLC. SLM: Spatial Light Modulator.} \label{Fig:schematic}
\end{figure*}

Unambiguous state discrimination was simulated for single-photon states of light encoded in high dimensions using single-outcome projective measurements~\cite{Agnew2014}.  Such projective measurements provide a limited functionality in quantum communication systems, where multi-outcome measurements are necessary for maximising key rates and loophole-free tests of Bell non-locality \cite{PhysRevLett.111.130406}.  Recent work on state discrimination includes sorting high-dimensional states using optimal measurement strategies \cite{solis2021enhanced}, multi-state quantum discrimination through optical networks \cite{laneve2022experimental}, and quantum state elimination \cite{Webb2022}.  Additionally, much of this recent work is related to the complementary field of classical deep optical networks used for information processing and image classification using diffractive optics \cite{lin2018, Wetzstein2020, kulce2021all}.  Methods for sorting and manipulating states using bulk optics \cite{Leach2002}, two phase screens \cite{Berkhout2010, Mirhosseini2013}, complex media \cite{defienne2016two,Leedumrongwatthanakun2020,goel2022inverse}, and multi-plane light converters (MPLCs)~\cite{Morizur2010,Labroille2014,Fontaine2017, Fontaine2019, Mounaix2020,Brandt2020,Hiekkamaki2020,Lib2021} have been the topic of significant recent research.


 The problem we set out to solve in this work is the simultaneous sorting of $d$ equally overlapping $d$-dimensional quantum states of light $\{\ket{\psi_1},...,\ket{\psi_d}\}$, i.e. with all pairwise fidelities of these states equal to each other $F = \left|\bra{\psi_i} \psi_{j \ne i}\rangle\right|^2$ \cite{franke-arnold12_unamb_state_discr_high_dimen}.  There are several requirements in order to perform this task. Firstly, every input mode in the set $\{\ket{\psi_1},...,\ket{\psi_d}\}$ is mapped to an individual measurement mode $\{ \ket{1},...,\ket{d} \}$ that uniquely identifies it, i.e.~when a photon in state $\ket{\psi_1}$ passes through the system, only the measurement mode $\ket{1}$ can ``click.''  The challenge is to perform this sorting process when every state in the set has a non-zero fidelity with respect to every other state in the set, i.e.~$0< \left|\braket{\psi_i | \psi_{j \ne i}}\right|^2 < 1$.  Secondly, the system should not make any errors and incorrectly identify any input state, i.e.~for the $\ket{\psi_1}$ input state, the probability of all output modes other than $\ket{1}$ clicking should be equal to zero.   Finally, the sorting process occurs simultaneously, and with the highest possible probability, which for USD is given by $\eta=1-|\bra{\psi_i} \psi_{j \ne i} \rangle|^2$ \cite{franke-arnold12_unamb_state_discr_high_dimen, Agnew2014}. 

\emph{MPLC for USD}:-- We realise unambiguous state discrimination (USD) within the framework of a multi-plane light converter (MPLC) as shown in Fig.~\ref{Fig:schematic}.  This allows us to simultaneously transform a set of non-orthogonal states in any basis into a new basis of spatially separated measurement modes that can be simultaneously detected by a position-resolving detector. Here, the non-orthogonal states of light we sort are super-positions of Hermite-Gaussian (HG) modes, and we design the output modes of the MPLC to be spatially separated Gaussian spots that can be directly read by a camera or a single-photon detector array.  The MPLC is programmed to perform the required unitary (USD operation) and the necessary mode conversion (HG $\rightarrow$ Gaussian spots) at the same time.

 The holograms used in MPLC devices are typically constructed using an inverse-design technique known as wavefront matching  \cite{Hashimoto2005,Sakamaki2007,Fontaine2017, Fontaine2019}. This is an iterative algorithm where the optical fields for the input and output modes are forward and backward propagated respectively and overlapped at each plane of the MPLC. The phase of the MPLC at each plane is calculated in such a way that the entire set of input modes are phase-matched to the respective output modes. This process is repeated for each plane sequentially until the algorithm converges and the difference between the forward and backwards propagating light is minimised.  Each reflection from a mask performs a phase-only transformation $\hat{\Phi}_i$ of the input states of light, which is followed by free-space propagation ${\hat{H}}$ to the next plane. The total operator of the device after $n$ reflections is given by $\hat{U} = \hat{H} \prod_{i=n}^1\left(\hat{\Phi}_i\hat{H} \right)$. The free-space propagation operator ${\hat{H}}$ is calculated by simulating light propagation in free-space as discussed in detail in Supplementary Information.

Here we introduce an additional output state labelled $\ket{?}$ into the wavefront matching protocol, which means that for $d$ input modes, there are now $d+1$ output modes. The MPLC device supports a large number of modes, which is ultimately limited by the number of pixels and spatial resolution of the phase masks. The number of modes of the MPLC greatly exceeds the number of modes we sort, which allows us to include the additional auxiliary mode in a straightforward manner.  We are free to choose $d$ input modes (superpositions of HG modes) and $d+1$ output modes (Gaussian spots placed symmetrically on the circumference of a circle). The wavefront matching technique then ensures the correct mapping between the two sets of modes.

The purpose of the $\ket{?}$ output is that if a photon is detected in this mode, it provides no information about the input state. However, this also implies that we have not made any incorrect identification, as required by USD.  As we can successfully discriminate between any two states with a probability of $1-F$, the probability that $\ket{?}$ clicks is equal to $F$. The key to the success of this protocol is that this operator is designed to perform both a unitary operation that maps a set of input states onto the required unambiguous measurement states and simultaneously changes the basis for the measurement outcomes to be spatially separated. Both of these transformations are performed concurrently within the MPLC.

\begin{figure} 
  \includegraphics[width=8.6cm]{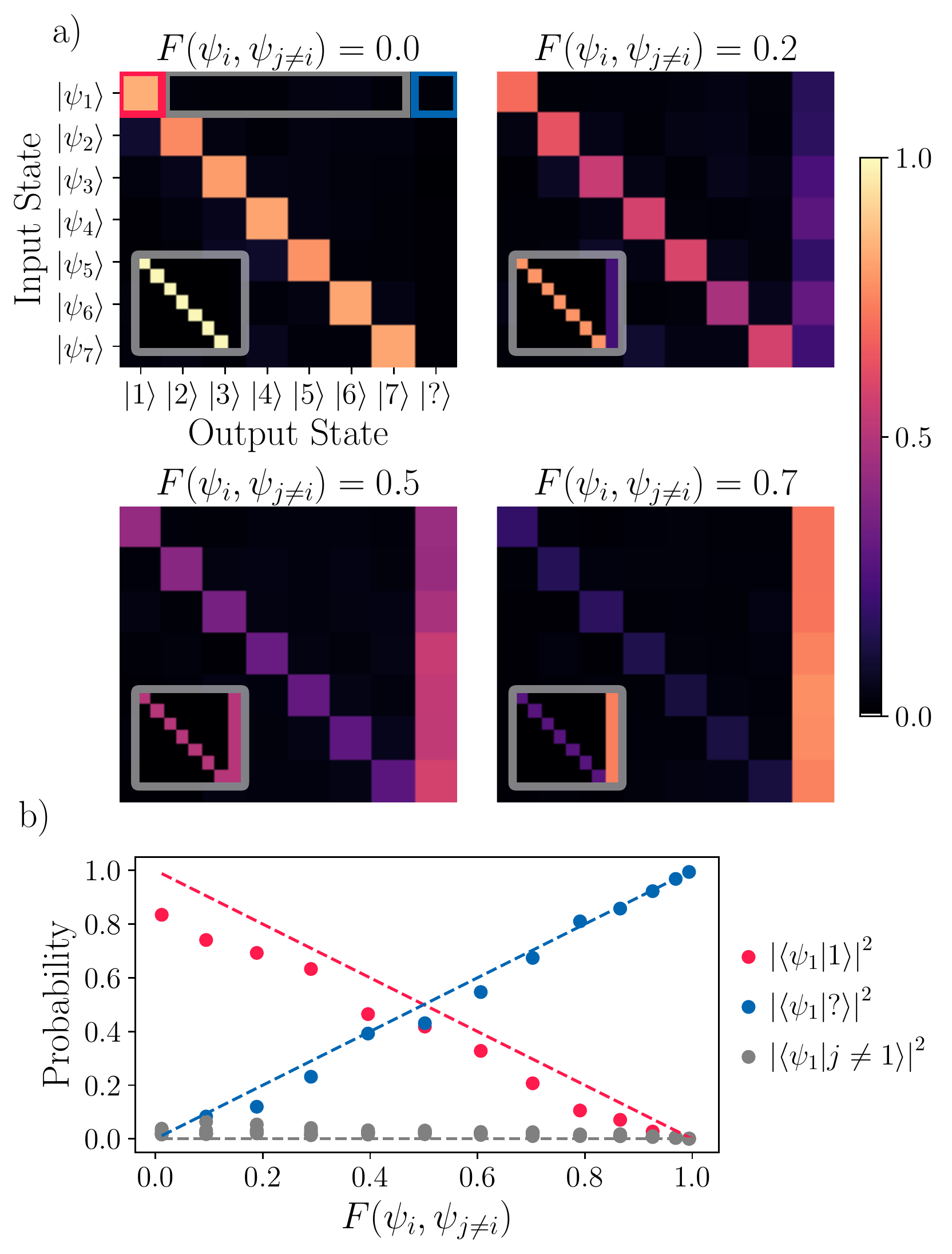}

\caption{Measurement data for simultaneous USD in $d = 7$ using an MPLC. (a) Measurement matrices for a range of fidelities ($F$) between the input states.  The insets show the numerically modelled results. The results are normalised by the total power detected in each output state. (b) Probability of measuring an input state $\ket{\psi_1}$ in a given output state $\ket{x}$ as a function of inter-state fidelity. This is a cross section of the measurement matrices for the $\ket{\psi_1}$ input state.  The points are the measured values, which should be compared to the theoretical predictions, indicated by the dashed lines.}
\label{fig:matrices_d7} 
\end{figure}

\begin{figure}
  \centering
  \includegraphics[width = 8.6cm]{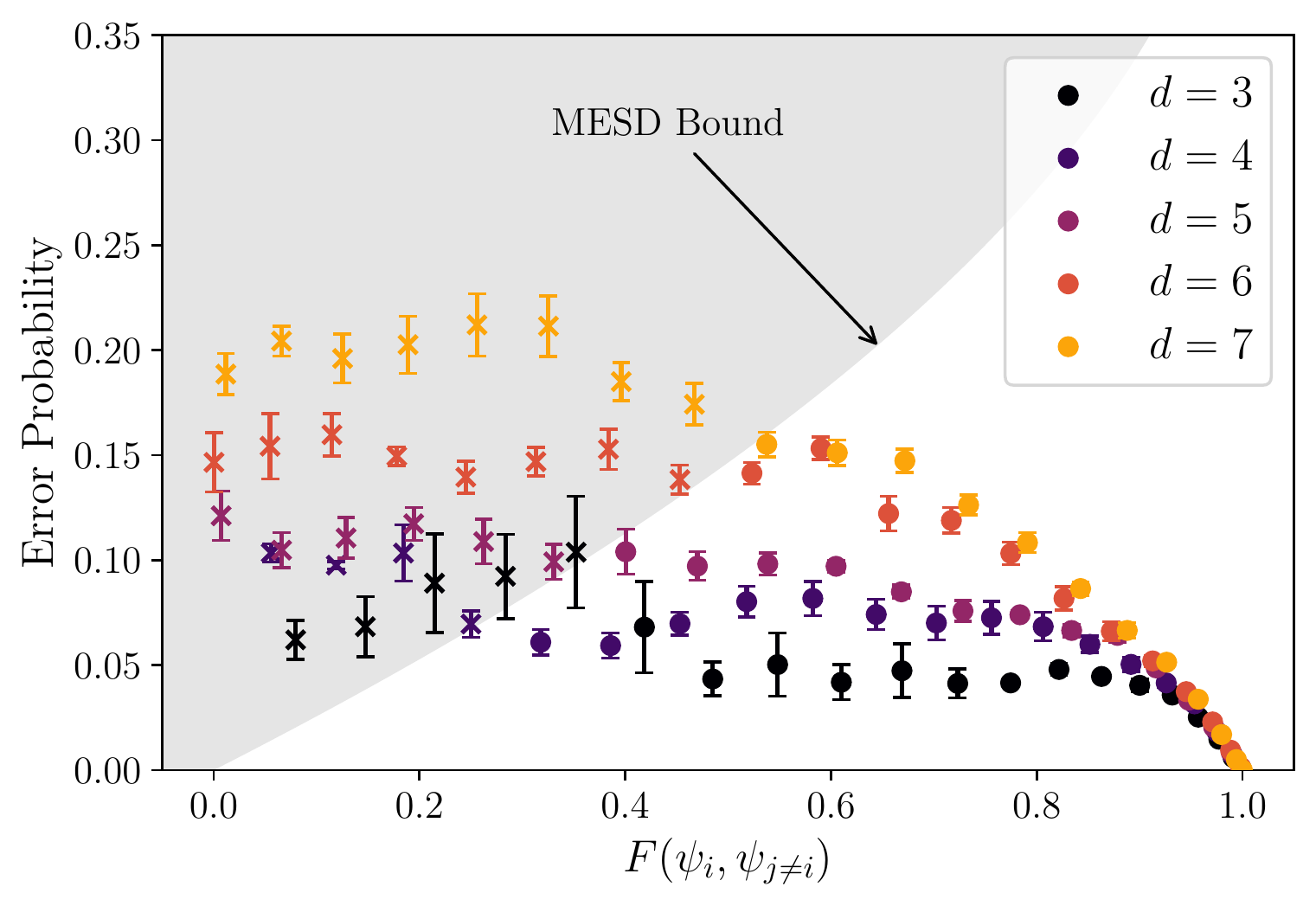}
  \caption{Evaluation of the MPLC for USD compared to the theoretical limit of MESD. The grey shaded area indicates the error-fidelity region accessible via MESD. Points outside this region represent USD with an error rate that is below than what is possible with MESD. }
\label{fig:evaluation}
\end{figure}

\emph{Results}:-- We performed unambiguous state discrimination for sets of symmetric non-orthogonal states constructed from modes in the Hermite-Gaussian (HG) basis.   In our experiment, we generated these modes with a HeNe laser and a spatial light modulator, and detected them with a CMOS camera.  

In each realisation of USD for high-dimensional states of light, $d$ non-orthogonal modes were transformed into $d+1$ output modes and measured simultaneously on a CMOS camera.  The intensities recorded by the camera pixels located in the output modes were integrated and converted to a detection probability.  We achieved USD for sets of states in dimensions ranging from $d = 3$ to $7$ for fidelities in the range $F(\psi_i, \psi_{j \ne i}) \in (0, 1)$. Each state was generated as a complex superposition of Hermite-Gauss modes while varying the inter-state fidelity from 0 to 1 (see Supplementary Information for further details).   The results for USD in 7 dimensions are displayed in the form of correlation matrices in Fig.~\ref{fig:matrices_d7}. Each row provides its detection probability in all possible outputs.   Fig.~\ref{fig:matrices_d7}b compares the experimentally measured probabilities of successful USD to the theoretical predictions. This data clearly demonstrates that unambiguous sorting is achieved for 7-dimensional overlapping states of light.

To quantitatively assess our system, we analyse the performance of the MPLC compared to the theoretical limit of minimum error state discrimination (MESD) \cite{PhysRevA.77.012328,Agnew2014}. No auxiliary state is used in MESD in $d$ dimensions, and the error is instead distributed between the $d$ output states. This leads to a probability of error ($p_{\mathrm{err}}$) in the output, which is defined as the probability of measuring any output state $\ket{j\neq i}$ when given an input state $\ket{\psi_i}$. MESD for uniform-fidelity states has a minimum possible error probability given by $p_{\mathrm{err}}\ge\frac{1}{2}\left(1-\sqrt{1-F(\psi_i, \psi_{j \ne i})}\right)$ \cite{PhysRevA.77.012328,Agnew2014}. In Fig.~\ref{fig:evaluation}, we plot the measured error probabilities of our USD protocol against this MESD threshold (grey area). We see that our system outperforms MESD and has a lower error rate than any strategy using MESD over a wide range of inter-state fidelities (overlaps). The error probabilities are higher for lower $F$ because the MPLC struggles more when sorting states that are farther apart. Additionally, as $d$ increases, so too does the total error. The performance reduction of the MPLC  for increasing $d$ can be explained as the accuracy of such transformations has been shown to reduce as the dimensionality increases for a fixed number of phase planes~\cite{goel2022inverse}.  Here, we use four phase plane while increasing the dimensionality of the set of modes that we are sorting.   We note that the error rate that we observe is approximately two times higher that that of Agnew \emph{et al.} \cite{Agnew2014}, where measurements were performed one outcome at a time, which necessarily includes $(d-1)/d$ amount of loss.  However, in this work, all outcomes were measured simultaneously, which is a significant advance over the prior work in terms of practical applications.

Theoretically, the error should be equal to zero for USD, yet we see the experimental implementation using the MPLC performs better for states with a higher initial fidelity (overlap).  The reason for this is that as $F$ increases, a higher fraction of the energy is put into the single ambiguous outcome.  This mapping, where all input states are sorted to a single output state, is a simpler task to achieve for the MPLC than the case where all input states are sorted to individual outcomes.    Additionally, we are comparing our experimental USD implementation to the theoretical limit of MESD.  We see that, consequently, there are is a region where the error probability does not fall below what MESD could achieve.  However, any practical implementation of high-dimensional MESD would be subject to similar error as our experiment and would not perform at this theoretical limit.

\begin{figure} 
  \centering

 \includegraphics[width = 8.6cm]{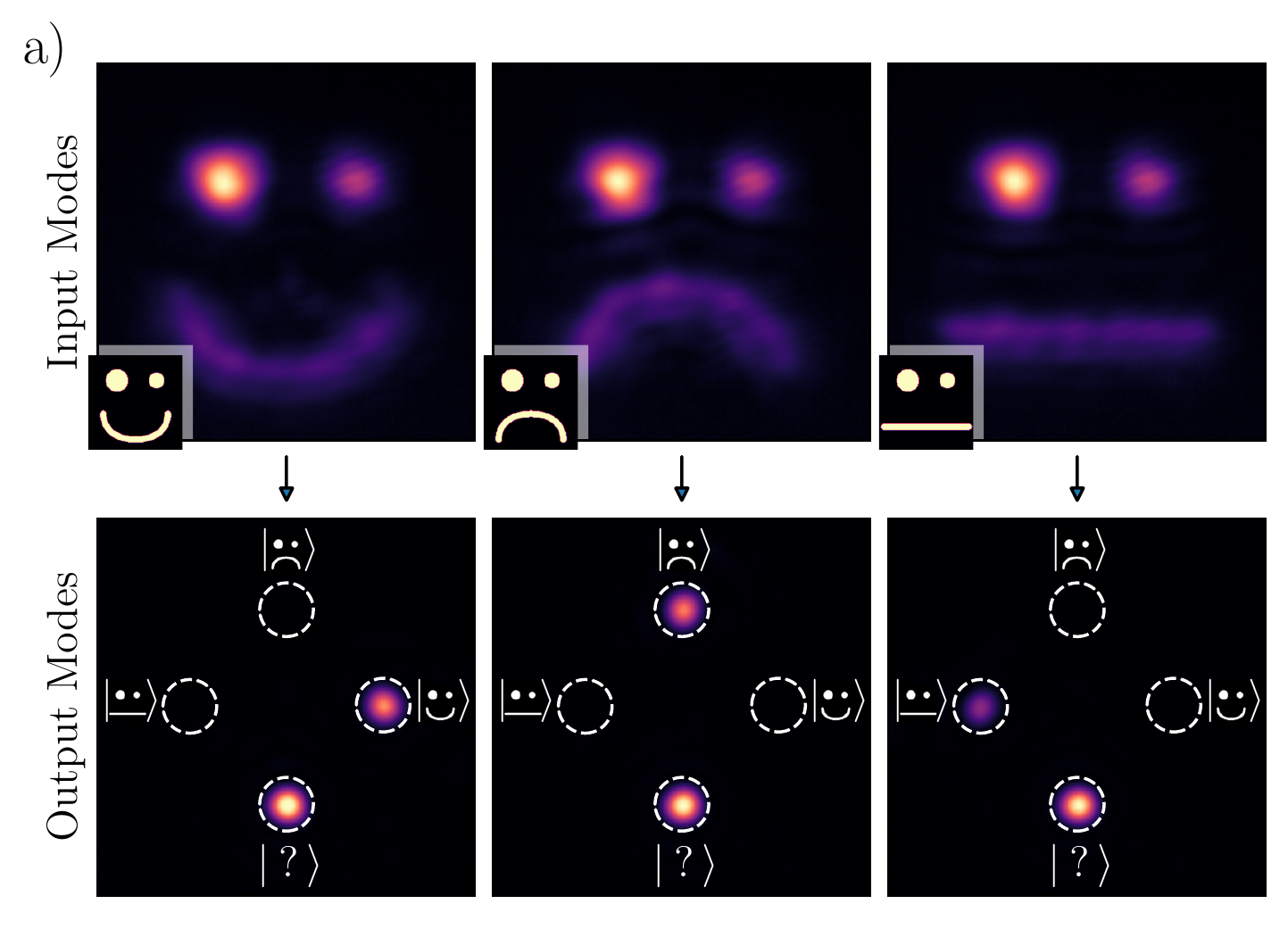}\\
  \includegraphics[width = 8.6cm]{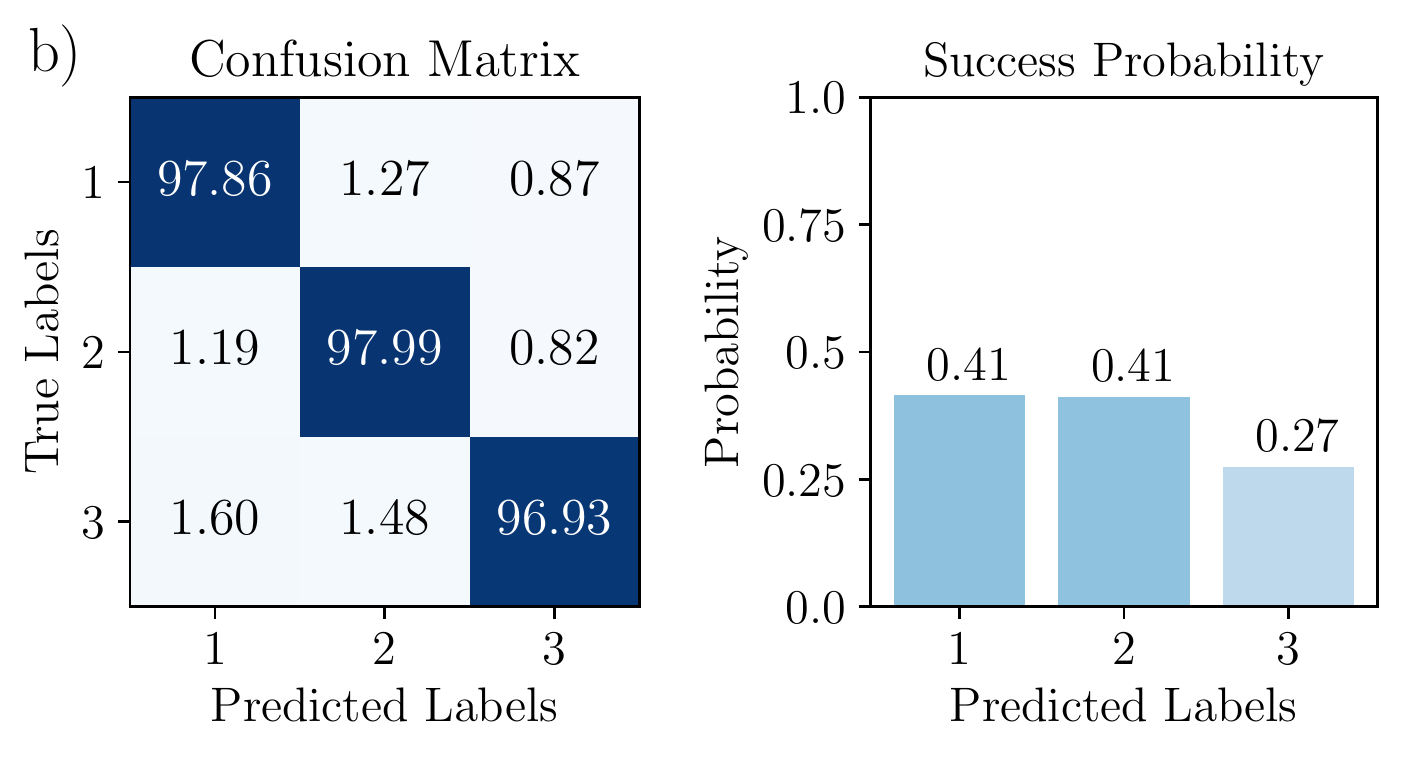}
\caption{Sorting overlapping images with the MPLC. The input images have a large fidelity ($F=0.34$) with each other, yet we can sort them with an accuracy of 97.6\%.  (a) Shows the measured intensities of the input and output modes. (b) Shows the confusion matrix of the sorting and the associated success probabilities. }
\label{fig:sorting_images}
\end{figure}

Figure \ref{fig:sorting_images} shows the extension of our method to the sorting of overlapping images.   We sort three images depicting a smiley face, sad face, and neutral face that have a large and symmetric overlap ($F=0.34$) with respect to each other---the eyes in the images are the same, while the mouth expressions are slightly different, connoting completely different emotions. In simple optical image classification, the three faces would be transformed directly to three spatially separated spots, and the wavefront matching algorithm would attempt to direct all input light into all output modes, leading to imperfect classification.  When using the extra mode $\ket{?}$, we have a place to direct any overlapping light, potentially leading to no errors in our measurement state outputs. 

 The correlation matrix for the images can be reformatted to a confusion matrix by removing the $\ket{?}$ mode and renormalising the rows. The success probability for image classification is then calculated as the average ratio of the light intensity in the outcomes of interest compared to the total light intensity of all other outcomes, excluding the ambiguous outcome.  Despite the large overlap between the input images, the USD protocol enables the sorting and classification with an average accuracy of 97.6\%. We see that the success probability is not constant across all three input states.  As numerical simulations suggest that an equal success probability can be obtained, we believe that this asymmetry is due to the combined experimental error of the generation of the input modes and the misalignment between phase masks  in the MPLC.

\emph{Conclusions}:-- In this work, we solve the measurement problem of simultaneous sorting of high-dimensional non-orthogonal states of light.  While previous works simulate a POVM with consecutive measurements, this work simultaneously realises the outcomes of a POVM  for high-dimensional non-orthogonal states of light, extending the use of MPLCs to include non-orthogonal input modes. Such simultaneous measurements can also be performed by engineering bases of light that are spatially separable on the detector~\cite{solis2017experimental}, however in this work we harness the abilities of MPLC to perform unitary operations in arbitrary spatial bases to perform this task experimentally for the first time.

The key to the success of this protocol is the additional output mode that provides extra flexibility to the system and enables perfect mapping from every $d$ non-orthogonal input state to a unique output. The method we adopt is the optimal strategy for minimising errors and correctly identifying the input states.  The consequence of USD is that at the single-photon level, the sorting does not provide an outcome 100\% of the time, and for intense modes of light, there is a reduction in the total power that is transmitted into the known output modes.  Future work will focus on extending our method to sort quantum states and modes that have a non-uniform fidelity with respect to each other, as well as applying such generalised measurement strategies on high-dimensional entangled states of light.

Although this experiment was performed with coherent states of light, the theoretical formalism still applies at the single-photon level since, in the case of linear optics, there is an equivalence between the probabilities associated with single-photon detection and classical laser fields \cite{Barnett_2022}.  The extension of our work, however, to include multi-photon states is of great interest, and in this case, multi-photon interference can lead to photon bunching or anti-bunching.  For multi-photon inputs, it would be necessary to measure the outcomes of our sorter in coincidence to fully reveal the photon statistics \cite{Hiekkamaki2020, defienne2016two}.

\emph{Acknowledgements}:--We thank Will McCutcheon for fruitful discussions regarding this work. This work was supported by EPSRC grants EP/T00097X/1 and EP/P024114/1, by QuantERA ERA-NET Co-fund (FWF Project I3773-N36), and the European Research Council (ERC) Starting grant PIQUaNT (950402).

\bibliography{references.bib}
\bibliographystyle{apsrev4-2.bst}






\newpage
\clearpage 
\renewcommand{\appendixpagename}{\begin{center}\large{\textbf{Supplementary Material : Simultaneously sorting overlapping quantum states of light}} \end{center}\label{SI}}

\onecolumngrid
\appendixpagename
\twocolumngrid

\setcounter{equation}{0}
\setcounter{figure}{0}
\setcounter{table}{0}
\renewcommand{\theequation}{S.\arabic{equation}}
\renewcommand{\thetable}{S.\arabic{table}}
\renewcommand{\thefigure}{S.\arabic{figure}}
\renewcommand{\theHfigure}{S.\arabic{figure}}




\newcommand{\psik}[1]{\ket{\psi_{#1}}}

\emph{Unambiguous State Discrimination}:-- In the process of unambiguous state discrimination (USD), our task is to accurately identify a random state from a set of states $S=
\left\{
  \psik{i}
\right\}_i$. In general, this set of states $S$ are non-orthogonal with each other. If two states have an overlap along a vector, when a state measurement collapses into this vector, we get no information of which state we started with. Since this measurement needs to be accurate, we denote this outcome as an ambiguous outcome. Alternatively if the result of the collapse is a vector which only has overlap with one of our initial states, then we can be sure that was the state we started with. The vectors we measure with are denoted $\ket{i}$ for the state which only has overlap with $\psik{i}$ and $\ket{?_j}$ for the state with multiple non-zero overlaps. This set of states forms a projective operator-valued measure (POVM). In general, there could be different overlaps between different states in $S$ corresponding to different ambiguous outcomes $\ket{?_j}$, however in this work we looked at symmetric states which have equal overlap with each other corresponding to a single ambiguous outcome denoted by $\ket{?}$.

\emph{Generation of symmetric states}:-- A set of $d$ $d$-dimensional symmetric states $\left\{\psik{i}\right\}_i$ is defined such that 
$\left|
  \braket{\psi_i | \psi_j}
\right|^2 = \delta_{ij} + (1-\delta_{ij})\left|\beta\right|^2$, where $\beta$ is the overlap between any two different states in the set. To be able to perform our USD measurements, we need a set of orthogonal measurement states ($\ket{i}$ and $\ket{?}$). To find these, we first construct the set of states $
\left\{
  \ket{i}
\right\}_i$ where $\ket{i}$ is orthogonal to every state but $\psik{i}$ ($\braket{i|\psi_j}=\delta_{ij}\alpha$). These states form an orthogonal basis for the initial set of states in $d$ dimensions. We then increase the dimension of the space to $d' = d+1$ by adding in an extra basis element $\ket{?}$ which is orthogonal to every $\ket{i}$. We now only need to find the transformation $\hat{V}$ which takes $\psik{i}$ ($= \sum_j\psi_{ij}\ket{j}=\alpha\ket{i}+\sum_{j\neq i}\psi_{ij}\ket{j}$) to $\ket{\psi_i'}=\hat{V}\psik{i}=\alpha\ket{i} + \beta \ket{?}$.

For this experiment, we constructed a set of states with a parameter $\theta$ which allowed us to control the fidelity. We start by building a set of $d$ states with overlap:

\begin{equation}
\braket{\psi'_i|\psi'_j} = -\frac{1}{d-1}
\end{equation}
These are generated with an iterative technique similar to Gram-Schmidt orthonormalisation. This theoretical technique is a generalization of the problem in three dimensions \cite{franke-arnold12_unamb_state_discr_high_dimen} to $d$-dimensions.
To generate the set of $d$ vectors $\left\{\ket{\psi_i}\right\}_i$ in $d-1$ dimensions:
Let $\psi_i^j$ be the $j$'th component of the $i$'th vector.

Find the components as:

\begin{equation}
  \psi_i^j = 
    \begin{cases}
        1, & \text{when }~i=j=1 \\
        \frac{-\frac{1}{d-1} - \sum\limits_{a < j}\psi_i^a\psi_j^{*a}}{\psi_j^j}, & \text{when } j < i \text{ or } j+1 = i=d \\
        \sqrt{1-\sum\limits_{a < j} \psi_i^a\psi_i^{*a}}, & \text{when } j = i \\
        0, & \text{otherwise}
    \end{cases}
\end{equation}

where $\ket{\psi'_i} = \sum_{j = 1}^{d} \ket{j}$.
Once these states are made, we mix them into the additional dimension by:

\begin{equation}
  \ket{\psi_i} = \sin(\theta)\ket{\psi'_i} + \cos(\theta)\ket{d+1}.
\end{equation}
%

Experimentally, the states $\{ \ket{1},\ket{2}...\ket{d+1}\}$ are chosen from the Hermite-Gauss basis. Thus each input state $\ket{\psi_i}$  is prepared as a superposition of $d+1$ Hermite-Gauss modes. For a given dimension of the extended Hilbert space $d+1$, we choose a certain set of  the Hermite-Gauss modes such that they follow $n+m+1=d+1$, where $m$ and $n$ are number of modes in either direction.


\emph{Sorting non-orthogonal states}:-- In order to perform USD measurements on these set of $d$ $d$-dimensional symmetric states $S = \left\{\psik{i}\right\}_i$, we need to define a set of measurement states. To do so, first calculate  $\ket{\psi_i^\bot}$ , which are orthogonal to every state but $\ket{\psi_i}$. This can be done by taking the SVD of all the $\ket{\psi_{j\neq i}}$ stacked as a matrix, and removing the components of $\ket{\psi}$ from the subspace this matrix spans. We then define the set of measurement states $\left\{D_i\right\}$ by expanding the Hilbert-space into a higher dimension using $\psik{d+1}$ which is orthogonal to every state in $S$ as 

\begin{equation}
  D_i = \ket{\psi_i^\bot} + \sqrt{-\braket{\psi_1^\bot|\psi_2^\bot}}\ket{d+1}
\end{equation}
and then normalise it. The overlap $\braket{\psi_i^\bot|\psi_{j\neq i}^\bot}=\braket{\psi_1^\bot|\psi_2^\bot}$ is the same for any two states $i\neq j$. The unknown state is included by Gram-Schmidt orthogonalisation with this set of states. We also need to extend our original states $\ket{\psi_i}$ into this space, and we do this just by adding the extra basis element $\ket{\psi_{d+1}}$ with $0$ as its coefficient.

The final sorting is achieved using multi-plane light conversion that converts the set of states $\left\{D_i\right\}$  into spatially separated outcomes on a camera.

\begin{figure}
  \centering
  \begin{tikzpicture}[scale=2]

   \tikzmath{\lighty = 0.7; \xwidth = 1;}
    \foreach \xoffset in {0, 1, 2, 3}
    {
        \tikzmath{\planename = \xoffset -1;}
     \begin{scope}[cm={0.4,0.3,0,1,(\xoffset * \xwidth,0)}]
       \node[transform shape] at (0.2,0.3) {\includegraphics[width=.04\textwidth]{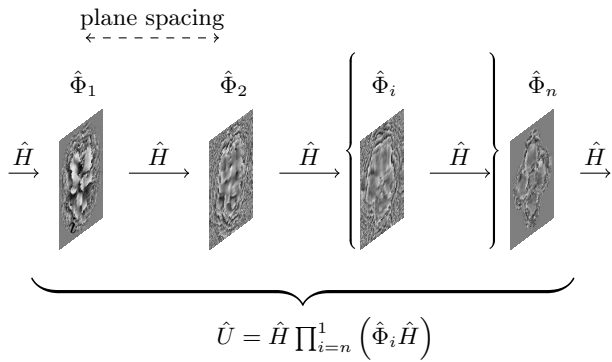}};
     \end{scope}
    }

      \begin{scope}[shift={(-1 * \xwidth, 0)}]
        \draw[->] (0.6, 0.4) -- (0.8, 0.4) node [midway, above] (TextNode) {$\hat{H}$};
       
     \end{scope}
    
    \foreach \xoffset in {1, 2}
    {
        \tikzmath{\planename = \xoffset -1;}
      \begin{scope}[shift={(\planename * \xwidth, 0)}]
        \draw[->] (0.4, 0.4) -- (0.8, 0.4) node [midway, above] (TextNode) {$\hat{H}$};
       \node at (0.1, 1) {$\hat{\Phi}_{\xoffset}$};

     \end{scope}
    }
      \begin{scope}[shift={(2, 0)}]
        \draw[->] (0.4, 0.4) -- (0.8, 0.4) node [midway, above] (TextNode) {$\hat{H}$};
       \node at (0.1, 1) {$\hat{\Phi}_{i}$};

      \end{scope}

       \begin{scope}[shift={(3 * \xwidth, 0)}]
        \draw[->] (0.4, 0.4) -- (0.6, 0.4) node [midway, above] (TextNode) {$\hat{H}$};
       
     \end{scope}

      \draw[decorate,thick,decoration={calligraphic brace,amplitude=3pt}] (1.9, -0.1) -- (1.9, 1.2);
      \draw[decorate,thick,decoration={calligraphic brace,amplitude=3pt}] (2.8, 1.2) -- (2.8, -0.1);
      node [end, right] (TextNode) {$i$};

      \begin{scope}[shift={(3, 0)}]
       \node at (0.1, 1) {$\hat{\Phi}_{n}$};
     \end{scope}

    \node at (0, 1.3) (nodeA){};
    \node at (1, 1.3) (nodeB){};
    \draw[<->,dashed] (nodeA) -- (nodeB) node [midway, above] (TextNode) {plane spacing};
    \draw [decorate,very thick,decoration={calligraphic brace,mirror,amplitude=10pt}] (-0.3, -0.3) -- (3.3, -0.3);
    
    \node at (1.65, -0.7) {$\hat{U} = 
      \hat{H}
        \prod_{i=n}^1\left(\hat{\Phi}_i\hat{H}
\right)$};
  \end{tikzpicture}
  \caption{ Illustrative diagram of MPLC: The unitary operation given as a product of multiple free-space propagation operators and phase planes.  }
  \label{fig:mplc_diagram}
\end{figure}

\emph{Multi Plane Light Conversion}:--  The multi-plane light converter (shown in Fig.~\ref{fig:mplc_diagram}) is a device which is trained to do a specific transformation on input light. It is formed of multiple planes in series which each impart a phase change on the light-field, and this phase change combined with the propagation of light between planes can approximate unitary transformations on the input light.

Plane $i$ performs a phase-only transformation $\hat{\Phi}_i$ to the light, which is followed by propagation to the next plane ${\hat{H}}$, giving the matrix of the device: $\hat{U} = 
      \hat{H}
        \prod_{i=n}^1\left(\hat{\Phi}_i\hat{H}
\right)$.

The calculation of the masks is done with the wavefront matching method.

\emph{Simulation of Light Propagation}:-- To simulate light propagation, we use the approximation of a 3-dimensional scalar field $\psi(\mathbf{x})$ \cite{siegman1986lasers,khare2015fourier}, and define an initial field over a plane $P$ orthogonal to the light's main propagation direction\ $\hat{\mathbf{n}}$: $
P = \left\{
  \mathbf{x} \mid (\mathbf{x} - \mathbf{a}) \cdot \hat{\mathbf{n}} = 0
\right\}$. The field is then propagated a distance $\alpha$ along the vector $\hat{\mathbf{n}}$ by the operator  $\hat{H}(\alpha)$

To approximate this calculation on the computer, we first discretise the field $\psi(\mathbf{x})$ over the plane P into the vector $\psi_{ij}$, where the indices $i$ and $j$ index two orthogonal directions. We take the Fourier transform of this vector in both of these indices to give $\tilde{\psi}=\mathcal{F}(\psi)$. The phase change over a distance $\alpha$ along the $\hat{\mathbf{n}}$ direction for light of spatial frequency $
\left|
  k
\right| = \frac{2\pi}{\lambda}$ is given by:

\begin{equation}
  \label{eq:phase_change}
 \exp
\left(
   -ik_{\hat{\mathbf{n}}}\alpha
 \right)
= \exp
\left(
  -i\sqrt{
    \left(
      \frac{2\pi}{\lambda}
\right)^2-
\left|
  k_{\bot\hat{\mathbf{n}}}
\right|^2}\alpha
\right)
\end{equation}
Where the value $\left|
  k_{\bot\hat{\mathbf{n}}}
\right|^2$ is sum of the squares of the coordinates $\tilde{\psi}$ is defined over. The inverse Fourier transform is then taken of this product to get the field at the new plane $\alpha\hat{\mathbf{n}}$ from the initial:
\begin{equation}
  \label{eq:light_simulation}
  \psi(\alpha) = \mathcal{F}^{-1}
  \left\{
     \exp
\left(
   -ik_{\hat{\mathbf{n}}}\alpha
 \right)\mathcal{F}
 \left(
   \psi_0
 \right)
  \right\}
\end{equation}

The propagation operator $\hat{H}(\alpha)$ is then given by: 

\begin{equation}
\label{eq:propagation_operator}
    \hat{H}(\alpha) = \hat{F_2}^\dagger \hat{D}(\alpha)\hat{F_2}
\end{equation}

 where $\hat{F_2}$ is the operator that takes 2D-Fourier transform of the field and  $\hat{D(\alpha)}$ is the operator that adds the distance dependent spatial phase to the field. 

\emph{Wavefront Matching}:--  In order to encode a transformation within an MPLC, we need to calculate the right patterns to be displayed on each plane. This is done by a process called ``wavefront matching''.  It consists of propagating the input modes forward through the planes, and the output modes backward through the planes. At the plane where the modes meet, we create a new mask as the average overlap over all modes. 

If our input modes at plane $p$ are $a_{i, p}(x)$, and our output modes are $b_{i, p}(x)$ then the expression for the next plane mask is given by:
\begin{widetext}

\begin{equation}
  \Phi_p'(x) = \sum_i a_{i,p}(x)b_{i,p}^*(x)\exp
  \left(
    -i\arg
    \left[
    \int dx \Phi_p^*(x) a_{i,p}(x)b_{i,p}^*(x)
    \right]
  \right)
\end{equation}

\end{widetext}
The forward, and backward propagating fields at plane $p$ are given by $a_{i,p}(x)$ and $b_{i,p}(x)$ respectively.
The goal is to have the fields exactly matching at those points, so that $\braket{a_{i,p}|\Phi_{p}^{\dagger}|b_{i,p}}=1$ over all input modes $i$.
The change in the phase mask can be written in terms of its phase:
\begin{align}
  \label{eq:new_phase_mask}
  \Phi'(x) &= e^{i(\theta(x) + \delta \theta(x))}  \\
  &\approx e^{i\theta(x)}
    \left(
    1+i\delta\theta(x)
    \right) \\
  &= \Phi(x)
    \left(
    1 + i\delta\theta(x)
    \right)
\end{align}

The new overlap is given by:
\begin{equation}
  \label{eq:new_overlap}
  \braket{a_{i,p}|\Phi'^{\dagger}|b_{i,p}}
\end{equation}

We want the new $\Phi'$ to increase the overlap:
\begin{widetext}

\begin{align}
  \label{eq:overlap}
  \eta' &= 
  \left|
  \braket{a_{i,p}|\Phi'^{\dagger}|b_{i,p}}
          \right|^{2} \\
  &\approx   \left|\braket{a_{i,p}|\Phi^{\dagger}(1-i\delta\theta^{\dagger})|b_{i,p}}
    \right|^{2} \\
  &=   \left|
  \braket{a_{i,p}|\Phi^{\dagger}|b_{i,p}} -i \braket{a_{i,p}|\Phi^{\dagger}\delta\theta^{\dagger}|b_{i,p}} 
    \right|^{2} \\
        &= \left(
  \braket{a_{i,p}|\Phi^{\dagger}|b_{i,p}} -i \braket{a_{i,p}|\Phi^{\dagger}\delta\theta^{\dagger}|b_{i,p}} 
          \right)
\overline{\left(
  \braket{a_{i,p}|\Phi^{\dagger}|b_{i,p}} -i \braket{a_{i,p}|\Phi^{\dagger}\delta\theta^{\dagger}|b_{i,p}} 
          \right)} \\
        &= \eta + 2\Re\left\{\
          i\braket{a_{i,p}|\Phi^{\dagger}|b_{i,p}}
\overline{\braket{a_{i,p}|\Phi^{\dagger}\delta\theta^{\dagger}|b_{i,p}} }
          \right\} + \mathcal{O}(\delta\theta^{2}) \\
  &\approx \eta -2\Im
    \left\{
    \braket{a_{i,p}|\Phi^{\dagger}|b_{i,p}}
\overline{\braket{a_{i,p}|\Phi^{\dagger}\delta\theta^{\dagger}|b_{i,p}} }
    \right\}
\end{align}

\end{widetext}
And so to satisfy $\eta'>\eta$, we just need:
\begin{equation}
  \label{eq:eta_greater}
  \Im
  \left\{
        \braket{a_{i,p}|\Phi^{\dagger}|b_{i,p}}
\overline{\braket{a_{i,p}|\Phi^{\dagger}\delta\theta^{\dagger}|b_{i,p}} }
  \right\} < 0
\end{equation}
We let $\alpha=\braket{a_{i,p}|\Phi^{\dagger}|b_{i,p}}$, and separate $\overline{\braket{a_{i,p}|\Phi^{\dagger}\delta\theta^{\dagger}|b_{i,p}} }$ into its components, requiring the inequality to be satisfied for each inidividual component:
\begin{equation}
  \label{eq:component_wise_inequality}
  \Im
  \left\{
    \alpha
    a_{i,p}(x)e^{i\theta(x)}\delta\theta(x)b_{i,p}^{\dagger}(x)
  \right\} < 0
\end{equation}
And letting $\gamma(x)=\alpha a_{i,p}(x)e^{i\theta(x)}b_{i,p}^{\dagger}(x)$:
\begin{equation}
  \label{eq:gamma_sub}
  \delta\theta(x)\Im
  \left\{
    \gamma(x)
  \right\}<0
\end{equation}
By letting $\delta\theta(x)=-\Im
\left\{
  \gamma(x)
\right\}$, we can satisfy this equation.
So, to update the phase mask:
\begin{equation}
  \label{eq:phase_mask_update}
\Phi'(x) = \sum_i e^{i\theta(x)} e^{
\left(
  -i\Im
\left\{
  \braket{a_{i,p}|\Phi^{\dagger}|b_{i,p}}
a_{i,p}(x)e^{i\theta(x)}b_{i,p}^{\dagger}(x)
\right\}
\right)}
\end{equation}

\emph{Addition of auxiliary mode}:-- Here we provide some explanation as to how the auxiliary mode is added to the system. There are an infinite number of free-space optical modes, but modes that are accessible in this experiment are limited to the Hilbert space of the MPLC.  This space is limited by the area and spatial frequencies of the masks on the SLM, which is set by the number of pixels on the SLM and the numerical aperture of the collection optics.  For example, if we want to sort a large number of input modes, the displayed masks on the SLM need to have a large range of spatial frequencies, which, in turn, requires many pixels.

It is the large number of modes provided by the MPLC enables the addition of the auxiliary mode.  We are free to choose where the outcomes of the MPLC are directed to, and we choose a spatially separated circular arrangement of spots.

In order to sort a greater number of mode modes, we would require larger masks and also larger spatial frequencies along with a greater number of phase planes.  However, there are physical constraints to this, for example, the propagation distance from one mask to the next, and the physical size the SLM that is used to display all the masks.   In our case, each mask is of dimension 250$\times$250 pixels, and we use four masks.  This enables us to sort up to seven non-orthogonal modes. For a comprehensive analysis of the trade-offs and scalability of multi-plane optical circuits in this context, see ref \cite{goel2022inverse}. 

\begin{figure*}
  \centering
\includegraphics[width=16cm]{  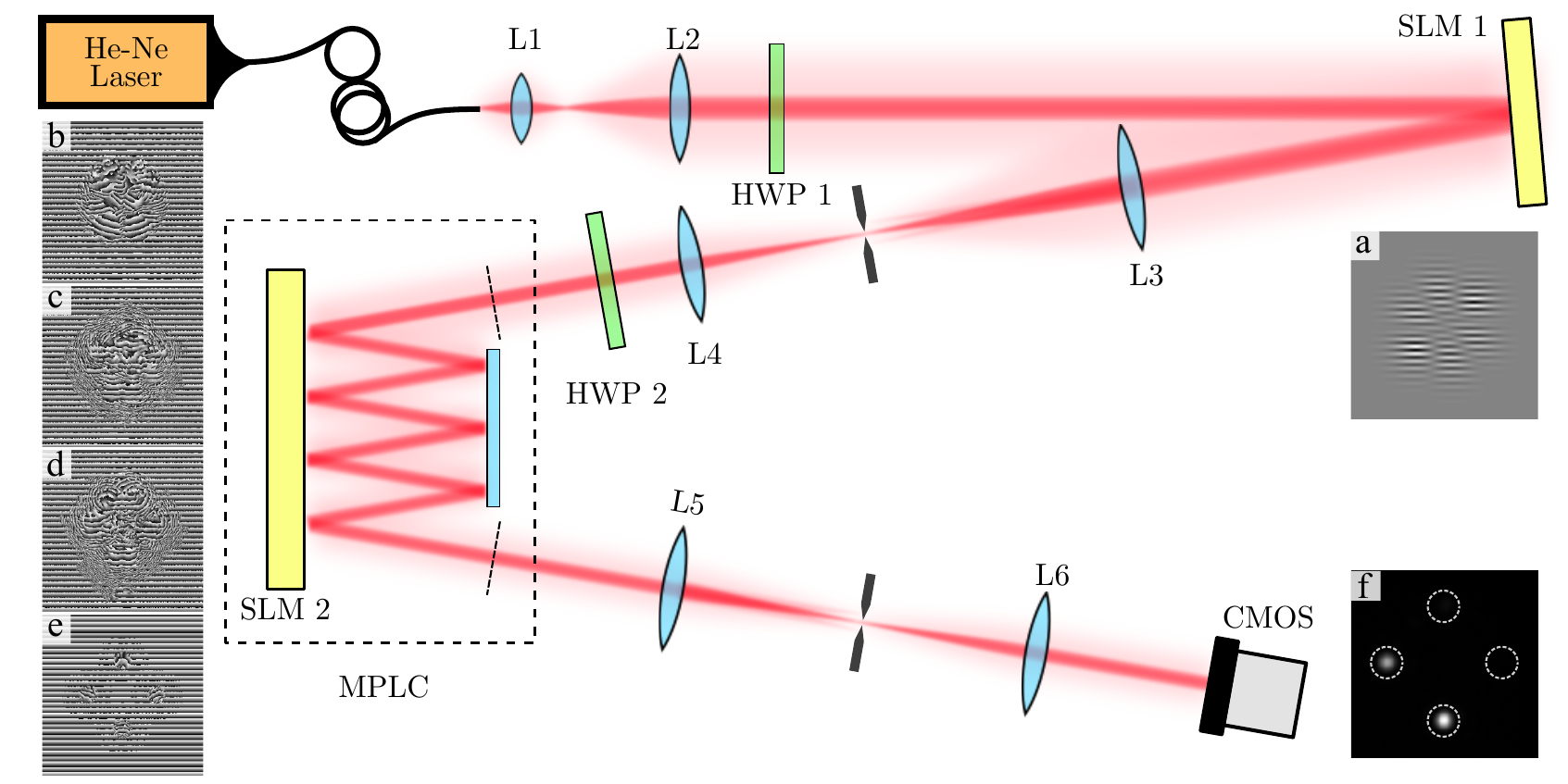}
  \caption{\textbf{Schematic of the experiment:} Input modes are generated using CGH holograms at SLM 1 (a).  The light from SLM 1 is filtered and imaged to the MPLC where 4 phase masks with a constant grating (b-e) are displayed. The output modes of the MPLC are filtered and imaged to a CMOS camera where light is sorted into Gaussian spots (f). Lenses L1-6 have focal length 45mm, 100mm, 750mm, 500mm, 250mm and 250mm, respectively. HWP 1 and HWP 2 are half-wave plates.}
  \label{fig:detailed_setup}
\end{figure*}

\emph{Experimental Setup}:-- 
As illustrated in  Fig.~\ref{fig:detailed_setup}, the experimental setup consists of three sections: generation, conversion and measurement. The desired non-orthogonal states are generated at a first SLM are then unambiguously discriminated at the second SLM, which acts as a multi-plane light converter (MPLC). The intensity distribution of the output states are then measured using a camera.  

For the generation of the input states, we first use a 5~mW He-Ne Laser to produce a linearly polarized beam at 633~nm. This is then passed though a single-mode fiber and is collimated with a pair of lenses with focal lengths 45~mm and 100~mm. This beam then passes through a half-wave plate (HWP) such that it is polarized in the optical axis of the generation SLM (Holoeye Pluto 2.1). Our input modes are generated using computer generated holograms (CGH), which are calculated using the type 3 method as detailed in \cite{Arrizon:05}. This beam is imaged to the input plane of the MPLC using two lenses of focal length 750~mm and 500~mm.  We use a Fourier filter at the focus of the 750~mm lens to select generated beams.

The MPLC is build in the reflection mode with 4 phase planes comprising of a phase-only SLM (Meadowlark E19x12) and a mirror, separated by 17 mm. Inside the MPLC, light undergo a transformation due to multiple reflection from the SLM followed by free-space propagation. Each reflection within the MPLC changes the phase profile of the mode. A grating of 10-pixel period is displayed across the whole SLM in order to efficiently separate modulated light from the residual. The target input and output modes are assigned at $17~$mm propagation distance before the first and after the last reflection from the SLM. 

Finally, the intensity distributions of the output modes are measured via the CMOS Camera (Thorlabs USB3.0) using a 4-$f$ system with a focal length of 250 mm. Since our beam is reflected though the same grating 4 times, at the focal plane of this imaging system, we filter the $4^th$ order diffraction that carries the fully transformed modes.

\emph{Data Processing}:-- We measure the experimental coupling matrices between the set of input modes and the output modes by measuring the intensity on each Gaussian spot. There are discrepancies between the ideal and measured intensities. This is due to a variety of factors like mode-dependent loss, imperfect wavefront matching, misalignments and subnormalised input state generation using computer-generated holograms (CGH).

We correct for these imperfections by multiplying the measured matrices with a ``correction vector''. Experimentally, this can be interpreted as applying a non-uniform attenuation to different ouput modes. If our ideal USD matrix is $M$, and our experimentally measured USD matrix is $E$, we can construct a correction vector $\mathbf{v}$ with elements:
\begin{equation}
  \label{eq:correction_vector_components}
  \mathbf{v} =
  \begin{pmatrix}
    \frac{E_{11}}{M_{11}} & \frac{E_{22}}{M_{22}} & \dots & \frac{E_{dd}}{M_{dd}} & \frac{\sum_{i=1}^d{E_{i(d+1)}}}{\sum_{i=1}^d{M_{i(d+1)}}}
  \end{pmatrix}
\end{equation}
Which can correct our measurements by multiplying column-wise:
\begin{equation}
  \label{eq:measurement_correction}
  M'_{ij} = M_{ij}\mathbf{v}_{j}
\end{equation}

\color{red}
\begin{figure*}
  \centering
\includegraphics[width=16cm]{  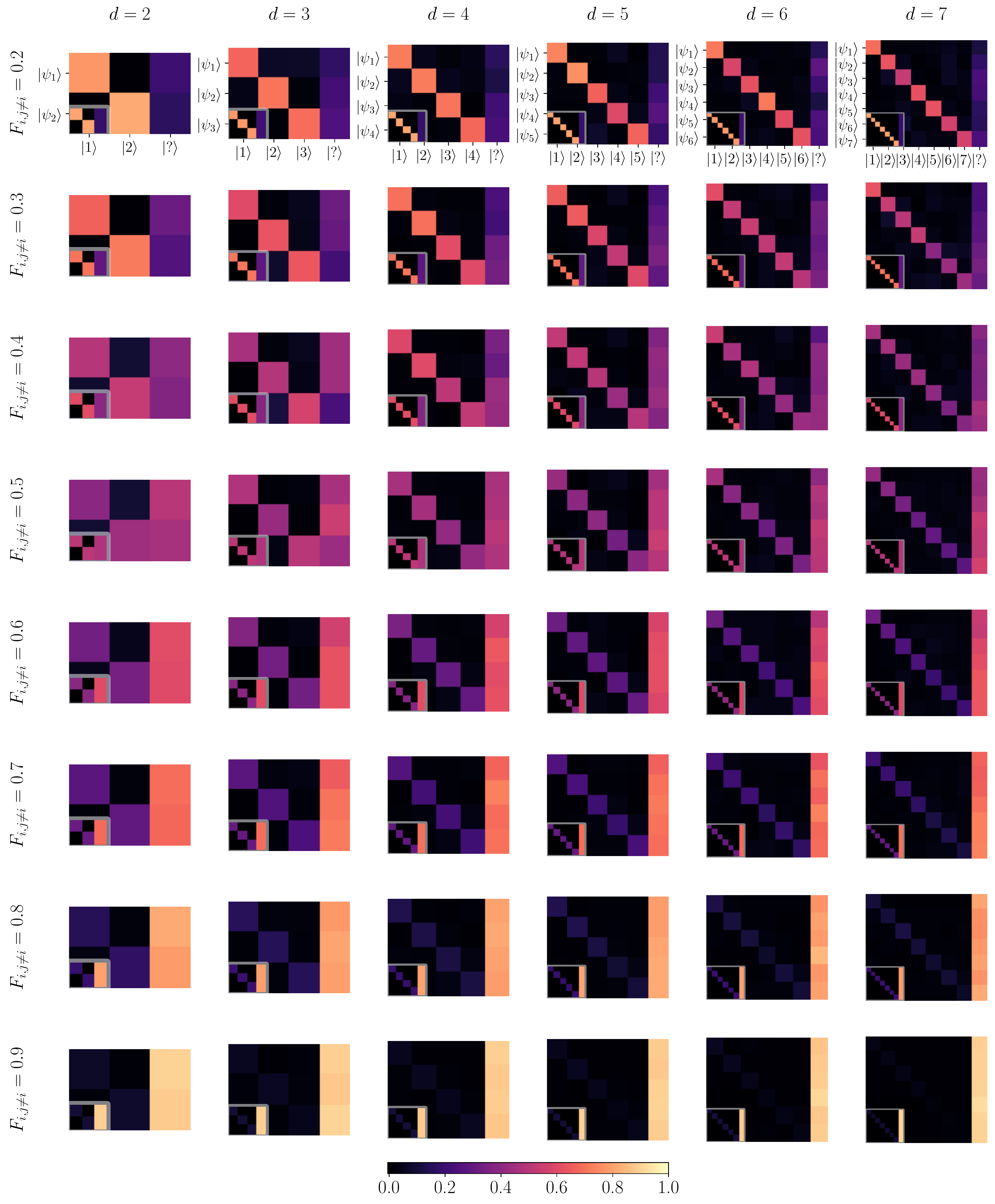}
  \caption{\textbf{Correlation matrices for all dimensions}: Correlation matrices of USD measurements in dimensions $d=2$ to $d=8$ for various input state fidelities ranging from $F=0.2$ to $F=0.9$}
  \label{fig:all_dim_data}
\end{figure*}

\begin{figure*}
  \centering
    \includegraphics[width=16cm]{  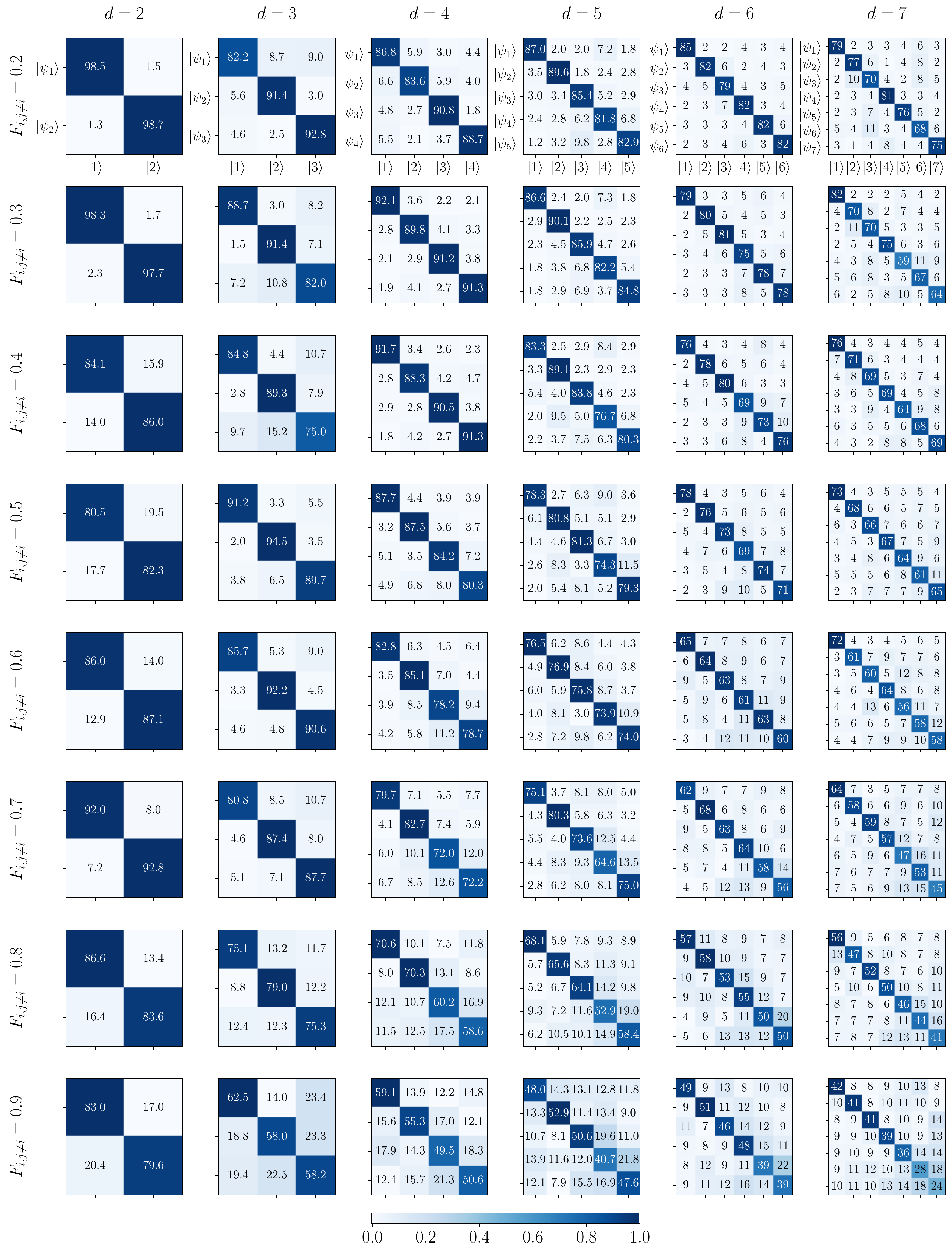}
  \caption{\textbf{Confusion matrices for all dimensions}: Confusion matrices of USD measurements in dimensions $d=2$ to $d=8$ for various input state fidelities ranging from $F=0.2$ to $F=0.9$ }
  \label{fig:all_dim_data}
\end{figure*}

\color{black}



\end{document}